# Stretchable optical diffraction grating from poly(acrylic acid)/polyethylene oxide stereocomplex

JINGHAN HE,[1,†] ANDRE KOVACH,[2,†] YUNXIANG WANG,[3] WILLIAM WANG,[2,4] WEI WU,[3] AND ANDREA M. ARMANI [1,2,3*]

[1]Department of Chemistry, University of Southern California, Los Angeles, CA 90089 USA
[2]Mork Family Department of Chemical Engineering and Materials Science, University of Southern California, Los Angeles, California 90089, USA
[3]Ming Hsieh Department of Electrical Engineering-Electrophysics, University of Southern California, Los Angeles, California 90089, USA
[4]Polytechnic School, Pasadena, California 91106, USA
*armani@usc.edu

**Abstract:** While initially static elements, advances in optical materials have enabled dynamically tunable optical diffraction gratings to be designed. One common tuning strategy relies on mechanically deformating of the grating pitch to modify the diffraction pattern. In the present work, we demonstrate an all-polymer tunable diffraction grating fabricated using a modified replica molding process. The poly(acrylic acid) (PAA)/polyethylene oxide (PEO) polymer stereocomplex films exhibit optical transmittance at or above 80% from 500 nm to 1400 nm and stretchability over 800% strain with reversibility under 70% strain. The imprinted gratings are characterized at 633 nm and 1064 nm under a range of strain conditions. The measured tunability agrees with the finite element method modeling.

Optomechanical systems are found throughout science and engineering, forming the foundation of microscopes, telescopes, and spectroscopy instrumentation as well as communication networks. Typically comprised of a complex array of lenses, gratings, and mirrors, these systems have been transformed from static to dynamically adaptable arrays through synergistic advances in materials chemistry and fabrication. While initial efforts focused on flexible mirrors for image correction in atmospheric science, the broader field of tunable or deformable optics has touched nearly all technical areas. One cross-cutting component is the diffraction grating.

Diffraction gratings are optical elements with a precisely defined periodic structure that can split or diffract an incident beam into a series of precisely spaced output beams [1]. Gratings can be designed to operate as either a transmissive or reflective structure, and the spacing of the diffracted beams is dependent on the incident angle, operating wavelength, grating pitch, and refractive index. Due to the flexibility of design and their ability to separate or spread optical wavelengths, diffraction gratings are used in numerous applications, including integrated optics, spectroscopy, and telecommunications [2–4]. As a result of this broad utility, the development of a reversible adaptive diffraction grating could have a broad impact.

The common strategy for designing a tunable grating relies on using electrical or thermal tuning of the optical properties of the material, but an emerging approach is based on strain or mechanical deformation of polymeric components. Unlike conventional crystalline materials, some polymers can exhibit large, reversible elastic response to mechanical strain [5–8]. Therefore, by applying tensile strain, any patterned optical structures will be deformed, leading to the tuning of the optical diffraction [9–12].

One advantage of using a polymer is its compatibility with non-conventional fabrication methods like replica molding and transfer molding. In these approaches, micro/nano patterns are initially fabricated using nanolithography and then replicated in the elastomer, or the micro/nanopattern is transferred to the polymer, which is used in lieu of a rigid substrate [13]. Replica molding specifically greatly simplifies the fabrication process, reducing the overall fabrication cost per part and accelerating manufacturing. Replica molding was first used to fabricate tunable diffraction gratings and other complex optical element topologies [14]. While successful, the tuning and optical transmission ranges were limited by the fundamental material performance of PDMS. Specifically, the common elongation at break of PDMS is less than 100% strain [15–18]. Therefore, to continue to push this field forward, the development of novel polymeric materials with high optical transmission and high mechanical stretchability (elongation at break) that are also compatible with molding or imprint methods is needed.

Recently, an optically transparent film of the hydrophilic poly(acrylic acid) (PAA)/polyethylene oxide (PEO) polymer stereocomplex has been reported with the highest strain of 1400% [19]. In addition, this material has demonstrated self-healing behavior based on dynamic intermolecular hydrogen bonds between COOH groups in PAA and oxygen groups in PEO. This property allows for an increased material longevity. However, to date, the development of a fabrication process to make self-supporting structures from this material has not been developed, limiting the applications that can be explored.

In this work, we develop and demonstrate an imprint molding inspired strategy to fabricate a deformable transmission diffraction grating based on the PAA/PEO stereocomplex. After characterizing the optical transparency and elasticity of the stereocomplex, the performance of a mechanically tunable grating is analyzed in the visible and near-IR. The diffraction pattern was reversibly changed by applying mechanical strain to the grating, and the results agree with simulation results.

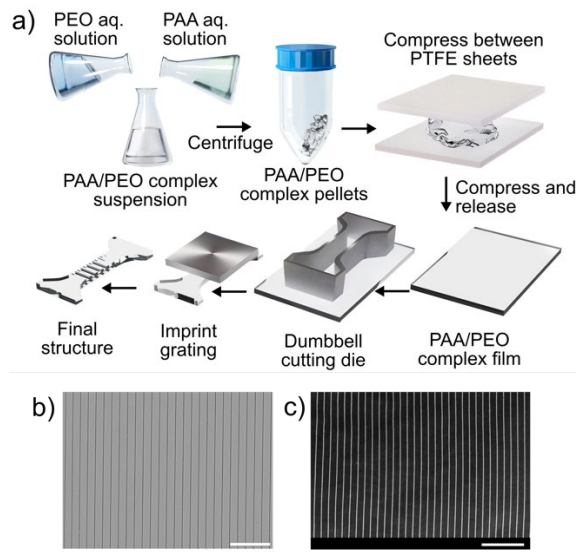

Fig. 1. (a) Outline of the material synthesis and the fabrication of the PAA/PEO polymer stereocomplex diffraction grating. SEM images of (b) Si master grating and (c) PAA/PEO polymer grating. Scale bar: 50 µm.

An overview of the fabrication process is shown in Fig. 1. To prepare a uniform, self-supporting polymer sample for optical and mechanical characterization, the PAA/PEO polymer stereocomplex pellets were first formed using procedures similar to previous works [8,20]. The molar ratio of COOH groups in PAA over oxygen segment in PEO is ideally 1:1 to facilitate the creation of the intermolecular hydrogen bonds, resulting in a uniform and robust network. To form a continuous, uniform, and self-supporting film (Fig. S1), the pellets were compressed between a pair of polytetrafluoroethylene (PTFE) sheets. The use of PTFE was particularly important to reduce stiction. Detailed methods on the pellet formation and film fabrication are included in the Supplementary Information. Before imprinting the grating, the basic optical and mechanical properties of the film are measured.

The optical transparency window of PAA/PEO polymer film was measured by a LAMBDA 950 UV-Vis Spectrophotometer (PerkinElmer). This polymer is ideally suited for optical applications as the transmission is at or above 80% from 500 nm to 1400 nm (Fig. S2). This high transmission in the near-IR range is particularly notable for a polymer. Subsequent diffraction measurements were performed at 633 nm and 1064 nm. At this pair of wavelengths, the transmission is 82% and 85%, respectively.

The PAA/PEO film was cut into a dumbbell shape using an ISO standard die cutter. The sample was mounted on an Instron tensile test instrument using a pair of pneumatic clamps to reduce sample damage (Fig. S3), and the mechanical response to tensile strain was analyzed (Fig. S4). As seen, the elongation at the break is above 800% strain, demonstrating significant elasticity of the film. However, there is also permanent deformation at these high strain values. To operate within the elastic region of the material resulting in minimal sample damage, less than 80% strain was used in subsequent grating measurements. In the cyclic testing of the polymer sample, the stereocomplex exhibits classic features of a visco-elastic material, or noticeable hysteretic behavior with significant energy loss after the first cycle. This is expected because of the dynamic hydrogen bonds in the PAA/PEO stereocomplex which can lead to strain accumulation and deformation of the sample. It should be noted that all five cycles were performed iteratively without any recovery time in between the measurements.

One unique feature of this material is its self-healing capability. Given sufficient time (~1 day), the material's mechanical response is recovered (Fig. S5). This self-healing property makes the material system more advantageous than conventional metal-based gratings or gratings based on stacked thin films. However, the hysteretic response is a concern. One strategy to tune or tailor this response would be improving the mechanical strength of the material with tunable physical or chemical structures to enhance the strain of the elastic region.

After completing the basic material analysis, the stretchable transmission grating was fabricated from a Si fabricated structure. The molding process was able to accurately replicate the nano and microscale features of the lithographically fabricated structure over a large area (Fig. 1 and S6). When performing the imprint step, the nanofabricated grating structure was oriented perpendicular to the vertical axis of the die cut-out and pressed into the polymer. The ISO standard die cutter is designed to ensure that the strain was uniformly distributed across the grating during deformation. As a result, the pitch and periodicity of the grating increased predictably according to the material's mechanical behavior (Fig. S5).

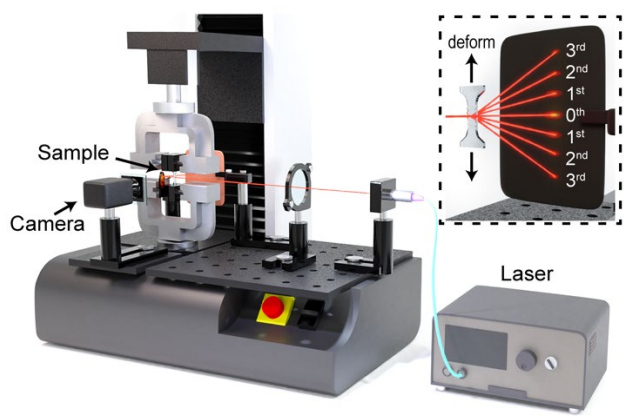

Fig. 2. Testing setup for the polymer grating consists of a 633 nm or 1064 nm laser, a universal tensile test machine (Instron) to apply continuous strains, a beam splitter, a receiving screen, and a camera to capture diffraction pattern images. Inset: Illustration of the grating structure and generated diffraction pattern. The diffraction orders (0th, 1st, 2nd, and 3rd) are indicated.

To characterize the optical performance of the stretchable grating, an optical test bench was integrated into the Instron (Fig. 2). The laser beam spot was focused on the grating with a collimator and a lens. A beam splitter, a receiving screen, and a camera were used to receive and detect the diffraction patterns. Data was taken at discrete strain values (steady state strain values) at 633 nm and at 1064 nm. Throughout all measurements, the incident angle of the laser beam and the separation distance between the grating and the detector card are constant. Thus, the separation distance between adjacent diffracted beams for the two wavelengths is a direct indicator of the dependence of the diffraction angle on wavelength.

A representative series of images from both wavelengths is shown in Fig. S7 and S8. A visual, qualitative inspection of the data reveals that separation distances, therefore the diffraction angles, at 1064 nm are larger than at 633 nm. As expected, at both wavelengths, the separation distance decreases with increasing strain, and then it recovers when the strain is removed.

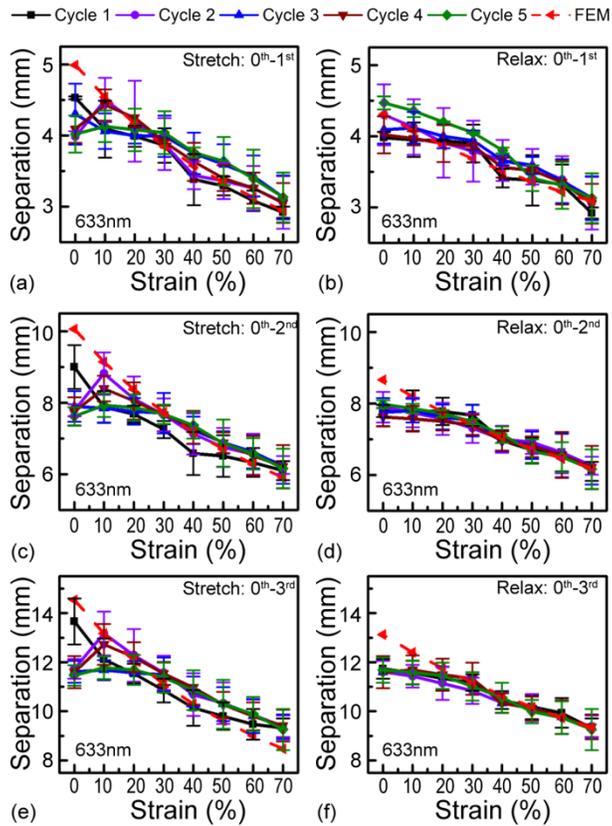

Fig. 3. Measurements and FEM simulations of strain-induced changes in the spacing between diffracted orders at 633 nm. Measurements are with respect to the 0th order. (a) 0th-1st order, stretching; (b) 0th-1st order, relaxing; (c) 0th-2nd order, stretching; (d) 0th-2nd order, relaxing; (e) 0th-3rd order, stretching; (f) 0th-3rd order, relaxing.

To analyze the data quantitatively, the diffraction order distances were determined using the scale on the receiving screen as a measure and analyzing the images using Python. Additionally, the grating was cyclically strained five times, mimicking the mechanical test measurement. The change in diffraction measured at 633 nm for all five cycles with respect to the 0th order is shown in Fig. 3. Additional analysis is located in Fig. S9. As the grating was stretched from 0% to 70% strain, the diffraction order distances of 0th-1st, 0th-2nd and 0th-3rd changed by ~1.5 mm, ~3 mm, and ~4.5 mm, respectively. During relaxation, these values systematically decreased by ~0.5-1 mm (Fig. 3b,d,f). These correspond to a change in diffraction angle of 0.9°, 2.4° and 3.5°, or absolute angle values of 3.8°, 7.5° and 11.2° at 0% strain. This nonlinear optical response can be attributed to the nonlinear visco-elastic mechanical behavior of the material (Fig. S5). However, it does give rise to several unique features worthy of discussion.

First, in Fig. 3a,c,e, for the second and almost all subsequent stretching cycles, the first strain step (0% to 10%) actually increases or has negligible impact on the diffraction order separation. Assuming a purely elastic material, this is the opposite behavior that would be expected. Additionally, the diffraction order separation does not fully return to the initial starting point after relaxation. This behavior was also observed in the mechanical test results. As discussed previously, most likely, both results are due to an accumulation of the structural damage in the material, also referred to as energy loss, which decreases the elasticity of the material. This hypothesis is supported by the material's ability to recover its mechanical behavior after resting for 24 hours.

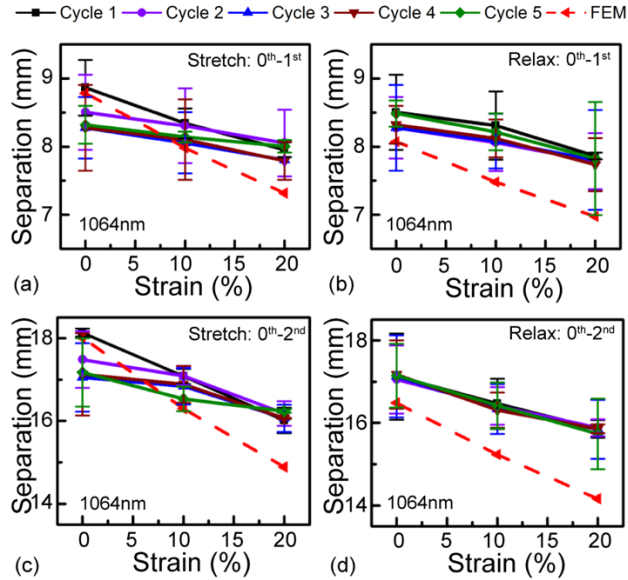

Fig. 4. Measurements and FEM simulations of strain-induced changes in the spacing between diffracted orders at 1064 nm. Measurements are with respect to the 0th order. (a) 0th-1st order, stretching; (b) 0th-1st order, relaxing; (c) 0th-2nd order, stretching; (d) 0th-2nd order, relaxing.

Fig. 4 summarizes the diffraction order distances measured at 1064 nm. Due to the significant increase in diffraction order spacing at 1064 nm, the imaging screen is only able to capture up to second order diffraction modes from 0% to 20% strains (Fig. S8(b)). As the grating was stretched from 0% to 20% strain, the diffraction order distances of 0th-1st and 0th-2nd changed by ~0.8 mm and ~2.0 mm, respectively. These correspond to a change in diffraction angle of 0.8° and 1.7°, or absolute diffraction angle values of 7.4° and 14.8° at 0% strain. Additionally, the diffraction efficiencies of each order/each wavelength at 0% strain are listed in Table S1. Similar to the findings at 633 nm, the accumulated strains in the polymer during cycling also limited the ability of the grating to fully recover. This was particularly pronounced during the first cycle.

To compare the experimental observations with theoretical predictions, the first three diffraction orders of the PAA/PEO polymer gratings were simulated using COMSOL Multiphysics, a finite element method (FEM) modeling software. The initial grating geometry was determined from scanning electron microscopy images of the experimental structure, and subsequent geometries were based on a combination of the measured elasticity of the material and the calculated deformation due to the applied strain. Additional details on the modeling are in the Supplementary Information. The diffraction was calculated at both visible (633 nm) and near-IR (1064 nm) incident wavelengths, and the results are plotted in the Fig. 3 and Fig. 4. As can be seen, the FEM calculations and the experimental findings are generally within experimental error at 633 and 1064 nm.

In summary, we have proposed and demonstrated a deformable polymer grating using a combination of nanoimprint and soft lithography. This modified replica molding approach provides a path for fabricating optical devices from this exciting class of stereocomplex polymers with self-healing properties. Using this method, we demonstrated an optical grating fabricated from a PAA/PEO polymer stereocomplex, characterizing fundamental optical and mechanical material properties and demonstrating its use as a diffraction grating. The fabrication method accurately transferred the micro-scale structures of the Si grating template into PAA/PEO polymer stereocomplex.

Unlike many transparent polymers, the PAA/PEO polymer stereocomplex has high optical transparency throughout most of the visible and near-IR wavelength ranges and exhibits a high percent elongation before failure (800%). Additionally, due to the intermolecular dynamic hydrogen bonds in the PAA/PEO stereocomplex, the polymer can be reversibly loaded and unloaded at lower strains in the elastic region at ~70% strain. Moreover, due to the self-healing behavior, given sufficient recovery time, the mechanical performance can be recovered. This combination of features makes it ideally suited for a deformable optical element. As a result, at both 633 nm and 1064 nm, the diffraction order distances agreed with the theoretical simulations.

This system opens novel pathways both in materials research and optics. As already mentioned, the development of improved materials systems will greatly advance this entire field, enabling a plethora of optical applications and devices, including tunable or flexible spectroscopy systems, wearable sensors and optical filters. Future studies may lie in the optimization of the material designs, integration of optically active components and fabrication of more complex optical elements.

**Funding.** Army Research Office (ARO) (W911NF1810033); Office of Naval Research (ONR) (N00014-17-1-2270, N00014-21-1-2048).

**Acknowledgments.** We thank Anna Shaposhnik (USC) for rendering. We thank Ruojiao Sun (USC), Raymond Yu (USC) and Rene Zeto (USC) for helpful discussions. We thank Core Center of Excellence in Nano Imaging (USC) for taking SEM images.

**Disclosures.** The authors declare no conflicts of interest.

**Data availability.** The data analysis code is freely available on GitHub. Data underlying the results presented in this paper are not publicly available at this time but may be obtained from the authors upon reasonable request.

[†]These authors contributed equally to this Letter.

**Supplemental document.** See Supplemental Information for supporting content.

## References

1. A. Guglielmelli, S. Nemati, A. E. Vasdekis, and L. D. Sio, J. Phys. D: Appl. Phys. 52, 053001 (2018).
2. P. Cheben, D.-X. Xu, S. Janz, and A. Densmore, Opt. Express 14, 4695 (2006).
3. M. A. Ettabib, Z. Liu, M. N. Zervas, and J. S. Wilkinson, Opt. Express 28, 37226 (2020).
4. G. Niederer, H. P. Herzig, J. Shamir, H. Thiele, M. Schnieper, and C. Zschokke, Appl. Opt. 43, 1683 (2004).
5. J. Y. Oh, S. Rondeau-Gagné, Y.-C. Chiu, A. Chortos, F. Lissel, G.-J. N. Wang, B. C. Schroeder, T. Kurosawa, J. Lopez, T. Katsumata, J. Xu, C. Zhu, X. Gu, W.-G. Bae, Y. Kim, L. Jin, J. W. Chung, J. B.-H. Tok, and Z. Bao, Nature 539, 411 (2016).
6. B. C. K. Tee and J. Ouyang, Adv. Mater. 30, 1802560 (2018).
7. K. Kim, Y.-G. Park, B. G. Hyun, M. Choi, and J.-U. Park, Adv. Mater. 31, 1804690 (2019).
8. Y. Wang, X. Liu, S. Li, T. Li, Y. Song, Z. Li, W. Zhang, and J. Sun, ACS Appl. Mater. Interfaces 9, 29120 (2017).
9. S. Geiger, J. Michon, S. Liu, J. Qin, J. Ni, J. Hu, T. Gu, and N. Lu, ACS Photonics 7, 2618 (2020).
10. A. N. Simonov, S. Grabarnik, and G. Vdovin, Opt. Express 15, 9784 (2007).
11. K. Yin, Y.-H. Lee, Z. He, and S.-T. Wu, Opt. Express 27, 5814 (2019).
12. W. Chen, W. Liu, Y. Jiang, M. Zhang, N. Song, N. J. Greybush, J. Guo, A. K. Estep, K. T. Turner, R. Agarwal, and C. R. Kagan, ACS Nano 12, 10683 (2018).
13. A. L. Martin, D. K. Armani, L. Yang, and K. J. Vahala, Opt. Lett. 29, 533 (2004).
14. J. A. Rogers, R. J. Jackman, O. J. A. Schueller, and G. M. Whitesides, Appl. Opt. 35, 6641 (1996).
15. D. Keskin, T. Mokabbar, Y. Pei, and P. Van Rijn, Polymers 10, 534 (2018).
16. K. M. Choi and J. A. Rogers, J. Am. Chem. Soc. 125, 4060 (2003).
17. N.-J. Huang, J. Zang, G.-D. Zhang, L.-Z. Guan, S.-N. Li, L. Zhao, and L.-C. Tang, RSC Adv. 7, 22045 (2017).
18. H. Fang, Y. Zhao, Y. Zhang, Y. Ren, and S.-L. Bai, ACS Appl. Mater. Interfaces 9, 26447 (2017).
19. W.-C. Liu, C.-H. Chung, and J.-L. Hong, ACS Omega 3, 11368 (2018).
20. Y. Wang, T. Li, S. Li, R. Guo, and J. Sun, ACS Appl. Mater. Interfaces 7, 13597 (2015).

**Full References**

1. A. Guglielmelli, S. Nemati, A. E. Vasdekis, and L. D. Sio, Stimuli responsive diffraction gratings in soft-composite materials. J. Phys. D: Appl. Phys. 52, 053001 (2018).
2. P. Cheben, D.-X. Xu, S. Janz, and A. Densmore, Subwavelength waveguide grating for mode conversion and light coupling in integrated optics. Opt. Express 14, 4695 (2006).
3. M. A. Ettabib, Z. Liu, M. N. Zervas, and J. S. Wilkinson, Optimized design for grating-coupled waveguide-enhanced Raman spectroscopy. Opt. Express 28, 37226 (2020).
4. G. Niederer, H. P. Herzig, J. Shamir, H. Thiele, M. Schnieper, and C. Zschokke, Tunable, oblique incidence resonant grating filter for telecommunications. Appl. Opt. 43, 1683 (2004).
5. J. Y. Oh, S. Rondeau-Gagné, Y.-C. Chiu, A. Chortos, F. Lissel, G.-J. N. Wang, B. C. Schroeder, T. Kurosawa, J. Lopez, T. Katsumata, J. Xu, C. Zhu, X. Gu, W.-G. Bae, Y. Kim, L. Jin, J. W. Chung, J. B.-H. Tok, and Z. Bao, Intrinsically stretchable and healable semiconducting polymer for organic transistors. Nature 539, 411 (2016).
6. B. C. K. Tee and J. Ouyang, Soft Electronically Functional Polymeric Composite Materials for a Flexible and Stretchable Digital Future. Adv. Mater. 30, 1802560 (2018).
7. K. Kim, Y.-G. Park, B. G. Hyun, M. Choi, and J.-U. Park, Recent Advances in Transparent Electronics with Stretchable Forms. Adv. Mater. 31, 1804690 (2019).
8. Y. Wang, X. Liu, S. Li, T. Li, Y. Song, Z. Li, W. Zhang, and J. Sun, Transparent, Healable Elastomers with High Mechanical Strength and Elasticity Derived from Hydrogen-Bonded Polymer Complexes. ACS Appl. Mater. Interfaces 9, 29120 (2017).
9. S. Geiger, J. Michon, S. Liu, J. Qin, J. Ni, J. Hu, T. Gu, and N. Lu, Flexible and Stretchable Photonics: The Next Stretch of Opportunities. ACS Photonics 7, 2618 (2020).
10. A. N. Simonov, S. Grabarnik, and G. Vdovin, Stretchable diffraction gratings for spectrometry. Opt. Express 15, 9784 (2007).
11. K. Yin, Y.-H. Lee, Z. He, and S.-T. Wu, Stretchable, flexible, rollable, and adherable polarization volume grating film. Opt. Express 27, 5814 (2019).
12. W. Chen, W. Liu, Y. Jiang, M. Zhang, N. Song, N. J. Greybush, J. Guo, A. K. Estep, K. T. Turner, R. Agarwal, and C. R. Kagan, Ultrasensitive, Mechanically Responsive Optical Metasurfaces via Strain Amplification. ACS Nano 12, 10683 (2018).
13. A. L. Martin, D. K. Armani, L. Yang, and K. J. Vahala, Replica-molded high-Q polymer microresonators. Opt. Lett. 29, 533 (2004).
14. J. A. Rogers, R. J. Jackman, O. J. A. Schueller, and G. M. Whitesides, Elastomeric diffraction gratings as photothermal detectors. Appl. Opt. 35, 6641 (1996).
15. D. Keskin, T. Mokabbar, Y. Pei, and P. Van Rijn, The Relationship between Bulk Silicone and Benzophenone-Initiated Hydrogel Coating Properties. Polymers 10, 534 (2018).
16. K. M. Choi and J. A. Rogers, A Photocurable Poly(dimethylsiloxane) Chemistry Designed for Soft Lithographic Molding and Printing in the Nanometer Regime. J. Am. Chem. Soc. 125, 4060 (2003).
17. N.-J. Huang, J. Zang, G.-D. Zhang, L.-Z. Guan, S.-N. Li, L. Zhao, and L.-C. Tang, Efficient interfacial interaction for improving mechanical properties of polydimethylsiloxane nanocomposites filled with low content of graphene oxide nanoribbons. RSC Adv. 7, 22045 (2017).
18. H. Fang, Y. Zhao, Y. Zhang, Y. Ren, and S.-L. Bai, Three-Dimensional Graphene Foam-Filled Elastomer Composites with High Thermal and Mechanical Properties. ACS Appl. Mater. Interfaces 9, 26447 (2017).
19. W.-C. Liu, C.-H. Chung, and J.-L. Hong, Highly Stretchable, Self-Healable Elastomers from Hydrogen-Bonded Interpolymer Complex (HIPC) and Their Use as Sensitive, Stable Electric Skin. ACS Omega 3, 11368 (2018).
20. Y. Wang, T. Li, S. Li, R. Guo, and J. Sun, Healable and Optically Transparent Polymeric Films Capable of Being Erased on Demand. ACS Appl. Mater. Interfaces 7, 13597 (2015).

# Stretchable optical diffraction grating from poly(acrylic acid)/polyethylene oxide stereocomplex

## 1. Simulations

The diffraction grating was modeled using the finite element method modeling (COMSOL Multiphysics; COMSOL, Inc.). The Young's modulus of PAA/PEO polymer is set to 4.1MPa based on our measurements. A single cell of the polymer grating is considered as a steady state where the dimension of the single cell is rearranged as strain is applied. The dimension of the cell and the size of the grating under each strain are obtained using solid mechanics simulation in COMSOL Multiphysics.

To simulate the optical properties of the PAA/PEO grating, a single cell was modeled using periodic boundary conditions in the in-plane dimensions and perfectly matched layers in the out-of-plane dimensions. PAA/PEO polymer stereocomplex is considered as the Cauchy model with Cauchy parameters A = 1.446, B = 0.01. The initial grating sizes, including the pitch, linewidth and height of the polymer grating, were set to 8.3 µm, 500 nm and 1 µm, respectively. When modeling the polymer grating at each strain, each single cell dimension and the size of the grating are redefined based on our mechanics simulation results.

## 2. Polymer film preparations

PAA ($M_w$ = 240 000, BeanTown Chemical) and PEO ($M_w$ = 1 000 000, Alfa Aesar) were used as received. The preparation protocol of PAA/PEO polymer stereocomplex was similar to previous works with a few differences noted here [1,2]. Typically, PEO aqueous solution (2.9 mg/mL, 50 mL, pH 2.8) and PAA aqueous solution (1.8 mg/mL, 50 mL, pH 2.8) were prepared in two separate beakers. Then, two aqueous solutions were slowly mixed into a third beaker via a syringe pump at a rate of 5 mL/min, precipitating PAA/PEO pellets which were further collected via centrifugation (7800 relative centrifugal force, overnight). To form a uniform polymer film (Fig. S1), PAA/PEO complex pellets were placed between two polytetrafluoroethylene (PTFE) sheets and compressed by Al blocks (~5 kPa) for one day before peeling off. A dumbbell-shaped film cutting die (Model Standard ISO 37-4) was used as the stamp to shape PAA/PEO film into a standard tensile specimen, forming a test sample for mechanical characterization of the basic polymer properties. Several thin films and dumbbell samples of this type were fabricated in order to allow the mechanical properties to be fully characterized.

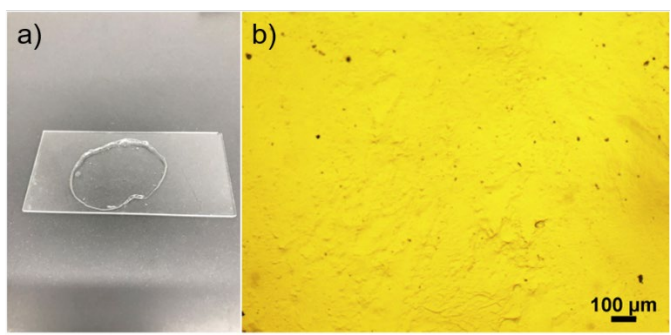

Fig. S1. (a) PAA/PEO polymer stereocomplex film on a piece of glass slide; (b) optical microscope image of PAA/PEO polymer stereocomplex.

### 3. UV-Vis study of PAA/PEO polymer film

The transmission of the polymer film was scanned using a LAMBDA 950 UV-Vis Spectrophotometer (PerkinElmer) where the polymer film is placed inside the spectrophotometer across the optical path to measure the transmittance at various wavelengths. The UV-Vis spectrum is shown in Fig. S2.

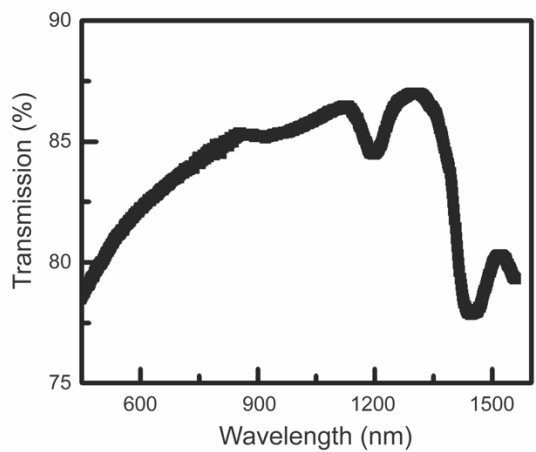

Fig. S2. UV-Vis spectrum of PAA/PEO stereocomplex film.

### 4. Fabrication of Si master grating

To begin with, a 100 nm lift-off underlayer (I-ULP 3.5% concentration, EZImprinting Inc.) was spin-coated on a Si wafer (WRS Materials) and baked at 120 °C for 5 minutes, which served as the adhesive layer. Then, another 100 nm UV nanoimprint resist (I-UVP 4.1% concentration, EZImprinting Inc.) layer was spin-coated onto the lift-off underlayer, and a subsequent nanoimprint lithography step was performed to form the grating pattern. Reactive ion etching (RIE, Oxford PlasmaPro 100) was performed to etch the residual UV imprint resist layer and the lift-off underlayer. Then, a 30 nm Cr layer was deposited by e-beam evaporation (Temescal BJD-1800 E-Beam Evaporator). To remove the lift-off underlayer and the layers above, a hot acetone bath was utilized, and the patterned Cr etching mask remained on the Si substrate, which was subsequently etched to 1 μm deep by RIE and removed the Cr etching mask. Lastly, to obtain the hydrophobic Si master grating, (heptadecafluoro-1,1,2,2-tetrahydrodecyl)trichlorosilane (GELEST Inc.) was deposited on the Si grating by chemical vapor deposition at room temperature for 30 min under the vacuum.

### 5. Fabrication of stretchable polymer grating

The Si master grating is molded into the center of a dumbbell shaped sample, forming the stretchable grating. Then, the hydrophobic Si master grating was placed upside down on top of the stamped polymer film to perform the replica molding (~5 kPa for 30 min at room temperature) of PAA/PEO grating. Finally, the hydrophobic Si master grating was peeled off to obtain the polymer grating. For initial inspection, a polymer grating was coated with platinum using sputter coating and imagined using scanning electron microscopy (FEI Nova NanoSEM).

## 6. Mechanical analysis

As shown in Fig. S3, the thin film, dumbbell shaped sample was mounted on the Instron 3340 using pneumatic clamps to minimize damage to the delicate polymer sample. Several mechanical tensile tests (including the cyclic loading-unloading tests) were performed to fully characterize the mechanical response of the material, including determining the breaking strength, hysteretic response, onset strain, and sample recovery. The stress-strain curve of PAA/PEO polymer stereocomplex (Fig. S4) indicates that the elastic region of the polymer is below ~70% strain. All measurements were performed at a stretching rate of 50% strain/min under ambient conditions, and samples were stored at ambient conditions when not being tested.

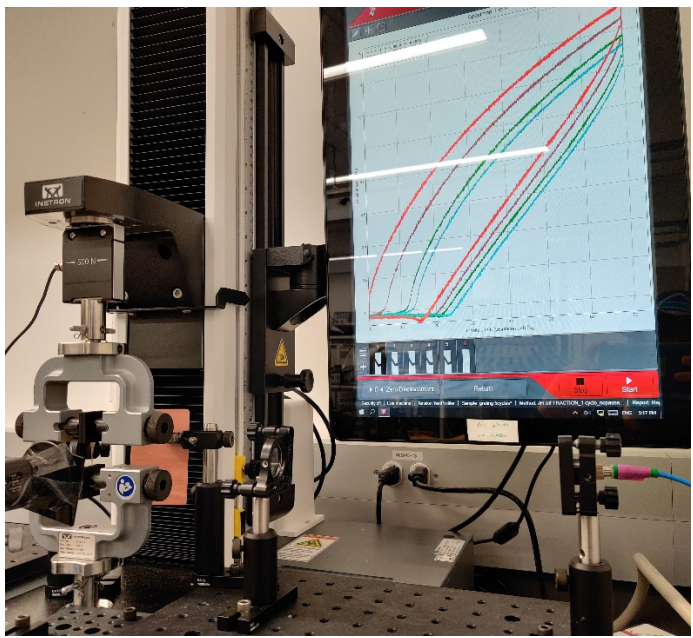

Fig. S3. Image of the actual optical testing setup.

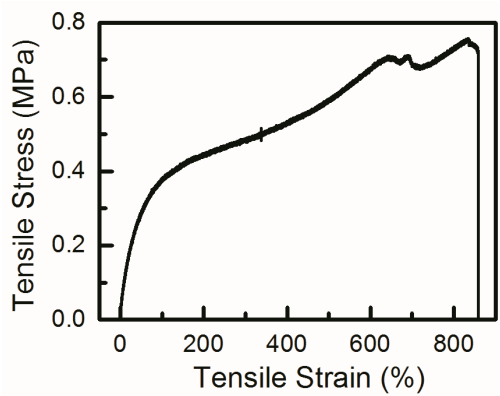

Fig. S4. Stress-strain curve of PAA/PEO polymer stereocomplex at the stretching rate of 50% min$^{-1}$.

The mechanical response of the PAA/PEO polymer to repeated load/unloading cycles is shown in Fig. S5(a). While there is a change in the polymer performance after the first cycle, the material is able to recover when it is allowed to relax for 24 hours (Fig. S5(b)). This recovery is directly related to the self-healing properties of the PAA/PEO stereocomplex. The onset strains shown in Fig. S5(c) and S5(d) are calculated from these data sets.

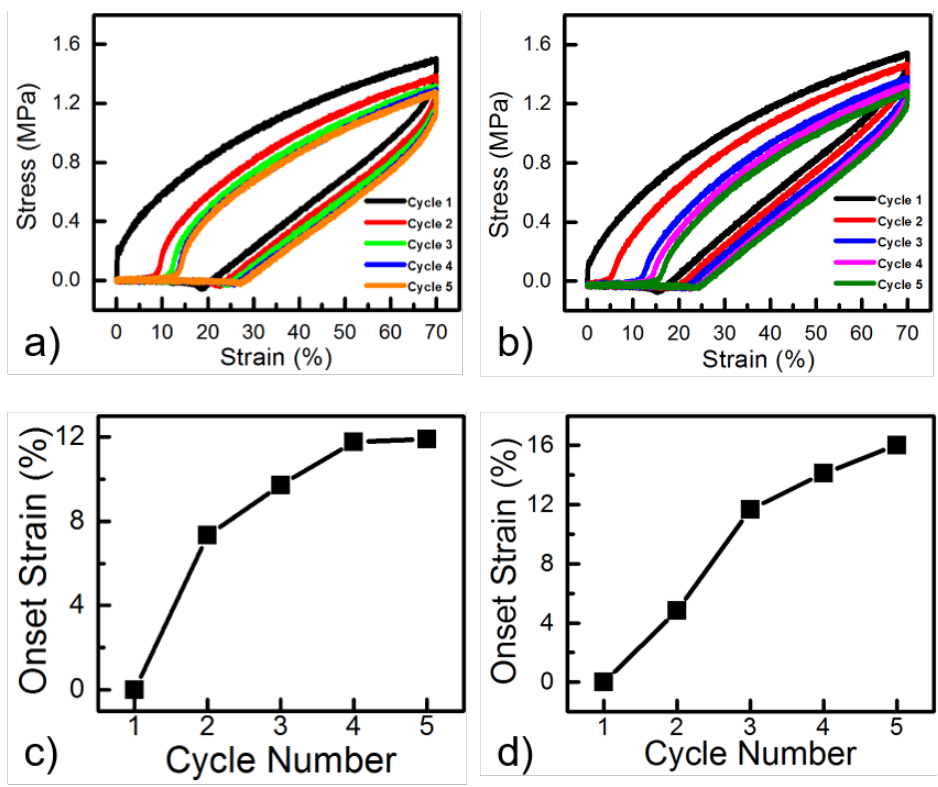

Fig. S5. The cyclic loading-unloading test results from (a) day 1 and (b) day 2, with one day relaxation of the polymer film between day 1 and day 2. The onset strain in each hysteresis loop cycle for (c) day 1 and (d) day 2. The cyclic curves of the PAA/PEO polymer film recovers after resting 1 day.

## 7. Scanning Electron Microscopy (SEM)

SEM images of the stretched and unstretched polymer gratings were acquired with a freshly made sample using a FEI Helios G4 P-FIB using a beam current of 1.6 nA and an accelerating voltage of 10 kV. Before imaging, a thin layer of platinum was sputter coated to reduce charging. Images are taken from the vertical direction and when the sample is tilted at 45°.

As seen in Fig. S6a, the initial periodicity of the grating at unstretched state (0% strain) is ~10 μm (Fig. S6a,b), and the surface has minor imperfections (Fig. S6c,d). When the polymer grating is stretched at 50% strain, which is the maximum possible inside the SEM, the periodicity of the grating is increased to ~15 μm (Fig. S6e, f), and the surface roughness or number of imperfections increases (Fig. S6g,h). This increase could be an explanation on the relatively low diffraction efficiency.

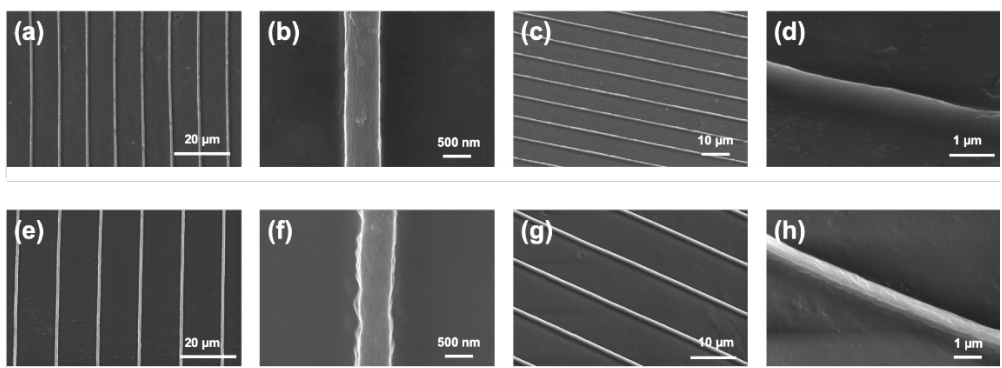

Fig. S6. SEM images of the PAA/PEO polymer grating. (a,b) Vertical direction view at 0% strain; (c,d) 45-degree tilted view at 0% strain; (e,f) vertical direction view at 50% strain; (g,h) 45-degree tilted view at 50% strain.

## 8. Optical characterization

To experimentally study how the PAA/PEO polymer grating is adaptively responded to the external mechanical strains, the strains on the polymer grating specimen were precisely controlled via the Instron universal tensile test instrument (Instron 3340, stretching rate 50% strain/min). There are several optical components integrated into the Instron system, as shown in Fig. 2.

Given the large optical transparency window, initial diffraction measurements using several lasers in the lab were performed (633 nm, 980 nm, 1064 nm, 1330 nm, 1550 nm), and the results are shown in Fig. S7. It is important to note that these initial exploratory measurements were performed with different lasers with different output powers. Due to the intrinsic laser type, the 633 nm and 1550 nm had the highest output power.

The laser beam spot was focused on the grating with a collimator and a lens. A beam splitter, a receiving screen (ThorLabs phosphor card) and a camera were used to detect and image the diffraction patterns. The phosphor card has an optimized response (or sensitivity) from ~800 nm-1200 nm. Therefore, when considering both laser power and detector card sensitivity, one would might expect a strong signal at 633 nm and 1550 nm (higher laser power) and 980 nm and 1064 nm (high sensitivity region of the card). The images were subsequently analyzed using a custom computer program Python, which has been uploaded to GitHub.

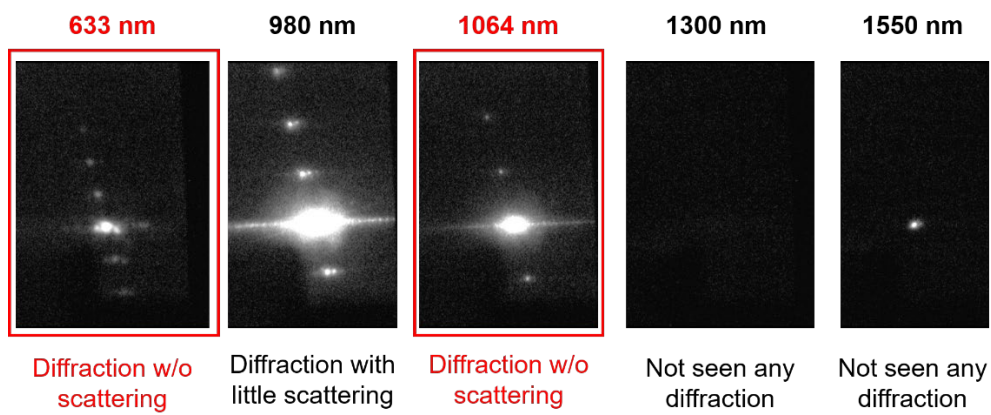

Fig. S7. The diffraction patterns of PAA/PEO polymer grating by various incident wavelengths. It turns out that 633 and 1064 nm are the best incident wavelengths because there are scatterings or no diffractions at other wavelengths due to our set-up with a specific IR card as the receiving screen and a specific beam splitter that works at a certain wavelength.

Diffraction can be easily observed at 633 nm, 980 nm, and 1064 nm (Fig. S7). No signal is apparent at 1300 nm, and the 0th order (or transmitted beam) can be seen at 1550 nm (Fig. S7). In general, these findings agree with what is expected based on the balance of laser power and detector card sensitivity. One reason for not detecting any diffraction at 1550 nm could be increased optical absorption of the material. Therefore, the present work focused on analyzing the diffraction at 633 nm and 1064 nm, to balance laser power, sensor card responsivity, and wavelength range. Example data sets at both wavelengths showing the change in diffraction pattern at different strain values (both stretch and relaxation) are shown in Fig. S8. The change in distance between diffraction orders as strain is applied for each cycle for 633 nm is shown in Fig. S9.

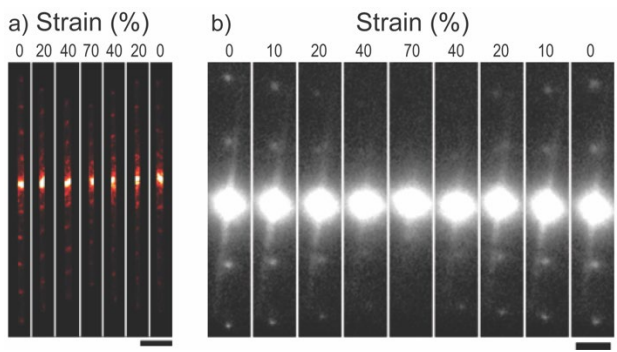

Fig. S8. The diffraction patterns generated at (a) 633 nm and (b) 1064 nm during one stretching-and-relaxing cycle are shown. The cycle started at 0% strain, increased to a maximum of 70% strain, and returned to 0% strain. Scale bar is 5 mm.

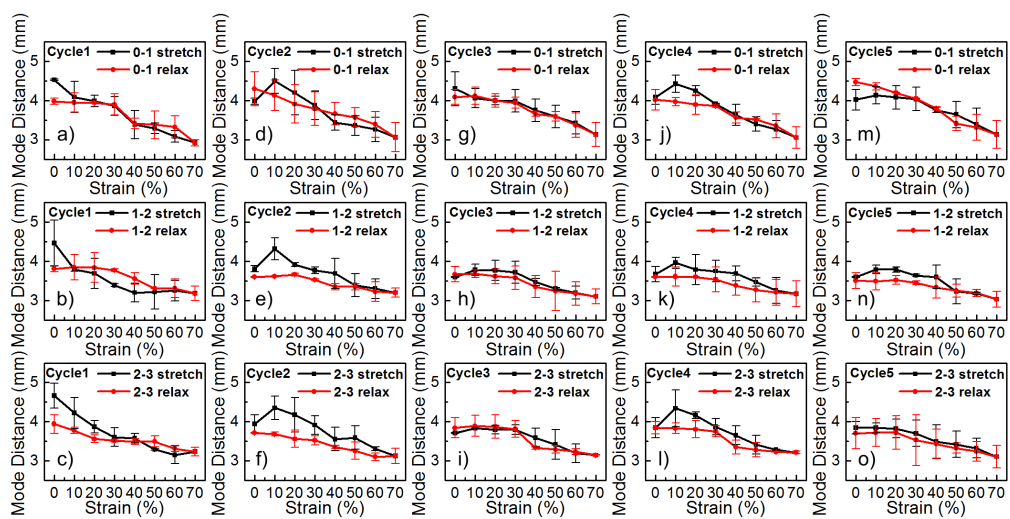

Fig. S9. Studies on the sequential diffraction order distances of the polymer grating at 633 nm in 5 individual cycles. (a,b,c) Cycle 1, (d,e,f) Cycle 2, (g,h,i) Cycle 3, (j,k,l) Cycle 4 and (m,n,o) Cycle 5 of the sequential diffraction order distances when stretching and relaxing.

The diffraction efficiencies of each order at 0% strain are listed in Table S1. The relative diffraction efficiencies were calculated with 0th-order as the benchmark.

**Table S1. The diffraction efficiencies of each order at various wavelengths.**

| Wavelength/nm | Diffraction Order | Relative Diffraction Efficiency/% |
| --- | --- | --- |
| 633 | 1st-order | 41 |
| 633 | 2nd-order | 15 |
| 633 | 3rd-order | 10 |
| 1064 | 1st-order | 29 |
| 1064 | 2nd-order | 18 |

## References

21. Y. Wang, X. Liu, S. Li, T. Li, Y. Song, Z. Li, W. Zhang, and J. Sun, ACS Appl. Mater. Interfaces 9, 29120 (2017).
22. Y. Wang, T. Li, S. Li, R. Guo, and J. Sun, ACS Appl. Mater. Interfaces 7, 13597 (2015).